\documentclass[aps,prl,twocolumn,superscriptaddress,showpacs,floatfix]{revtex4}
\usepackage{graphicx}
\usepackage{amssymb}
\usepackage{wasysym}
\usepackage{setspace}

\begin{document}

\title{Deterministic Photon Pairs via Coherent Optical Control of a Single Quantum Dot}

\author{Harishankar Jayakumar}
\email[]{harishankar@uibk.ac.at}
\affiliation{Institut f\"ur Experimentalphysik, Universit\"at Innsbruck, Technikerstrasse 25, 6020 Innsbruck, Austria}
\author{Ana Predojevi\'{c}}
\email[]{ana.predojevic@uibk.ac.at}
\affiliation{Institut f\"ur Experimentalphysik, Universit\"at Innsbruck, Technikerstrasse 25, 6020 Innsbruck, Austria}
\author{Tobias Huber}
\affiliation{Institut f\"ur Experimentalphysik, Universit\"at Innsbruck, Technikerstrasse 25, 6020 Innsbruck, Austria}
\author{Thomas Kauten}
\affiliation{Institut f\"ur Experimentalphysik, Universit\"at Innsbruck, Technikerstrasse 25, 6020 Innsbruck, Austria}
\author{Glenn S. Solomon}
\affiliation{Joint Quantum Institute, National Institute of Standards and Technology  \& University of Maryland, Gaithersburg, MD 20849, USA}
\author{Gregor Weihs}
\affiliation{Institut f\"ur Experimentalphysik, Universit\"at Innsbruck, Technikerstrasse 25, 6020 Innsbruck, Austria}

\begin{abstract}
The strong confinement of semiconductor excitons in a quantum dot gives rise to atom-like behavior. The full benefit of such a structure is best observed in resonant excitation where the excited state can be deterministically populated and coherently manipulated. Due to large refractive index and device geometry it remains challenging to observe resonantly excited emission that is free from laser scattering in III/V self-assembled quantum dots. Here we exploit the biexciton binding energy to create an extremely clean single photon source via two-photon resonant excitation of an InAs/GaAs quantum dot. We observe complete suppression of the excitation laser and multi-photon emissions. Additionally, we perform full coherent control of the ground-biexciton state qubit and observe an extended coherence time using an all-optical echo technique. The deterministic coherent photon pair creation makes this system suitable for the generation of time-bin entanglement and experiments on the interaction of photons from dissimilar sources.
\end{abstract}

\pacs{81.07.Ta, 78.55.Cr, 76.60.Lz}
\maketitle

Deterministic single photon sources exhibit the property of emitting one and only one photon with a very high probability at a desired time. Non-classical light sources like single photon and entangled photon pair sources are needed for linear optical quantum computation \cite{Knill}, long-distance quantum communication \cite{Duan} and protocols like teleportation \cite{Boschi} or entanglement swapping \cite{Pan}. Such sources have been demonstrated in stimulated emission of cavity-coupled atoms \cite{Hennrich} or heralded down-conversion sources \cite{Wolfgramm,Kwiat}. Quantum dots are proven sources of single photons, cascaded, and entangled \cite{Stevenson} photon pairs. To be used for quantum information processing, resonant excitation \cite{Ulhaq, Englund} and coherent manipulation \cite{Ramsey} are essential.

We present results obtained through resonant two-photon excitation \cite{Boyle, Flissikowski, Stufler} of the biexciton state of a single InAs/GaAs quantum dot embedded in a micro-cavity. In particular, we used pulsed laser light (4 ps) to coherently excite a single quantum dot using the lateral wave-guiding mode of a planar micro-cavity \cite{Muller} [Fig.~1(b)]. The two-photon excitation was performed via a virtual level half way in energy between the exciton and biexciton [Fig.~1(a)]. Owing to the difficulty of separating the excitation laser light from the emission, so far two-photon excitation of quantum dots has only been shown in other systems \cite{Boyle, Flissikowski, Miyazawa}, which are not useful as sources of high quality single photons. In contrast, in the present work, through a combination of techniques we were able to collect single photons and measure their statistics with high efficiency and virtually no background as shown in Fig.~2. 

\begin{figure*}[!ht]
\includegraphics[width=\textwidth]{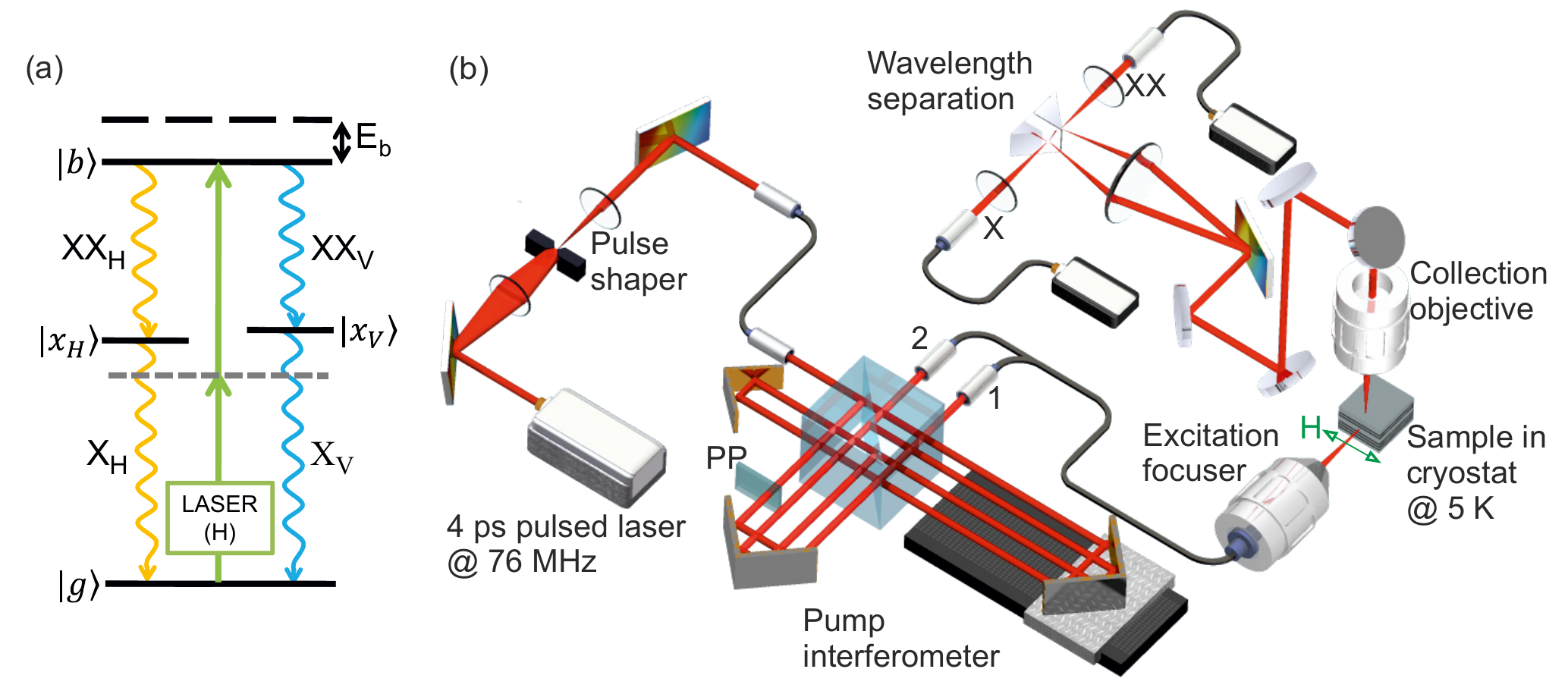}
\caption{ Excitation - emission scheme and experimental setup. (a) Energy level scheme for two-photon excitation of a biexciton and cascaded emission in a quantum dot. A pulsed laser with energy $E_{exc} = (| b\rangle-|g\rangle)/2$ and linearly polarized (H) along the cleaved edge of the sample, coherently couples the ground ($ |g\rangle$) and the biexciton ($|b\rangle$) states through a virtual level (dashed gray line). Biexciton recombination takes place through the intermediate exciton states ($|x_{H,V} \rangle$) emitting biexciton ($XX_{H,V}$) and exciton ($X_{H,V}$) photons. Biexciton binding energy ($\Delta_{b}$) results in ($|x_{H,V} \rangle-|g\rangle) > E_{exc} > ( |b\rangle-|x_{H,V} \rangle$). (b) Experimental set-up: consists of excitation, collection, and detection part.  Pump interferometer output fiber couplers labeled 1 and 2 collect double pulses and triple pulses respectively, for coherent control experiments. A glass phase plate on the short arm of the interferometer controls the intensity of the echo pulse. The spectral emission lines of interest are separated on a home-built grating spectrometer and coupled into single mode fibers.}
\end{figure*}

The emission probability of a resonantly driven system shows an oscillatory behavior as a function of the excitation pulse area known as Rabi oscillations. In our measurements we confirm the two-photon resonant excitation and the coherent nature of the excitation of the two-level system by observing an oscillation in emission intensity as a function of laser intensity \cite{Stievater, Zrenner, Wang} [Fig.~3(a)]. In the graphs the emission intensities are all normalized by the same numerical factor to the theoretically predicted values of the emission probability (see supplementary information), which are plotted as solid lines. 

The Rabi oscillations result from an exchange of the population between the ground state and the biexciton state; however, the limited number of oscillations in Fig.~3(a) is the result of decoherence.  Any process that transfers population to another state will destroy the coherence in the population exchange and therefore damp the Rabi oscillations. An obvious possibility could be the spontaneous radiative decay from the biexciton to the exciton level whose lifetime we have measured to be 405 ps. Nevertheless we are not affected by this kind of dephasing because we use laser pulses that are two orders of magnitude shorter.

A second possibility is the damping of the Rabi oscillations due to the proximity of the virtual level to the exciton state (detuned by $\Delta_e$= 335 GHz or 1 meV). The gray theory line in Fig.~3(a) shows that this is a very minor effect. A third possible dephasing process could be based on interaction with lattice phonons whose energies ($k_{B}T$ $\sim$ 400 $\mu$eV) could transfer the population from the virtual two-photon resonance to the exciton state  \cite{Machnikowski}. However, at sufficiently high detuning this process should cease due to insufficient energy of the lattice phonons.

\begin{figure}[!ht]
\includegraphics[width=0.95\linewidth]{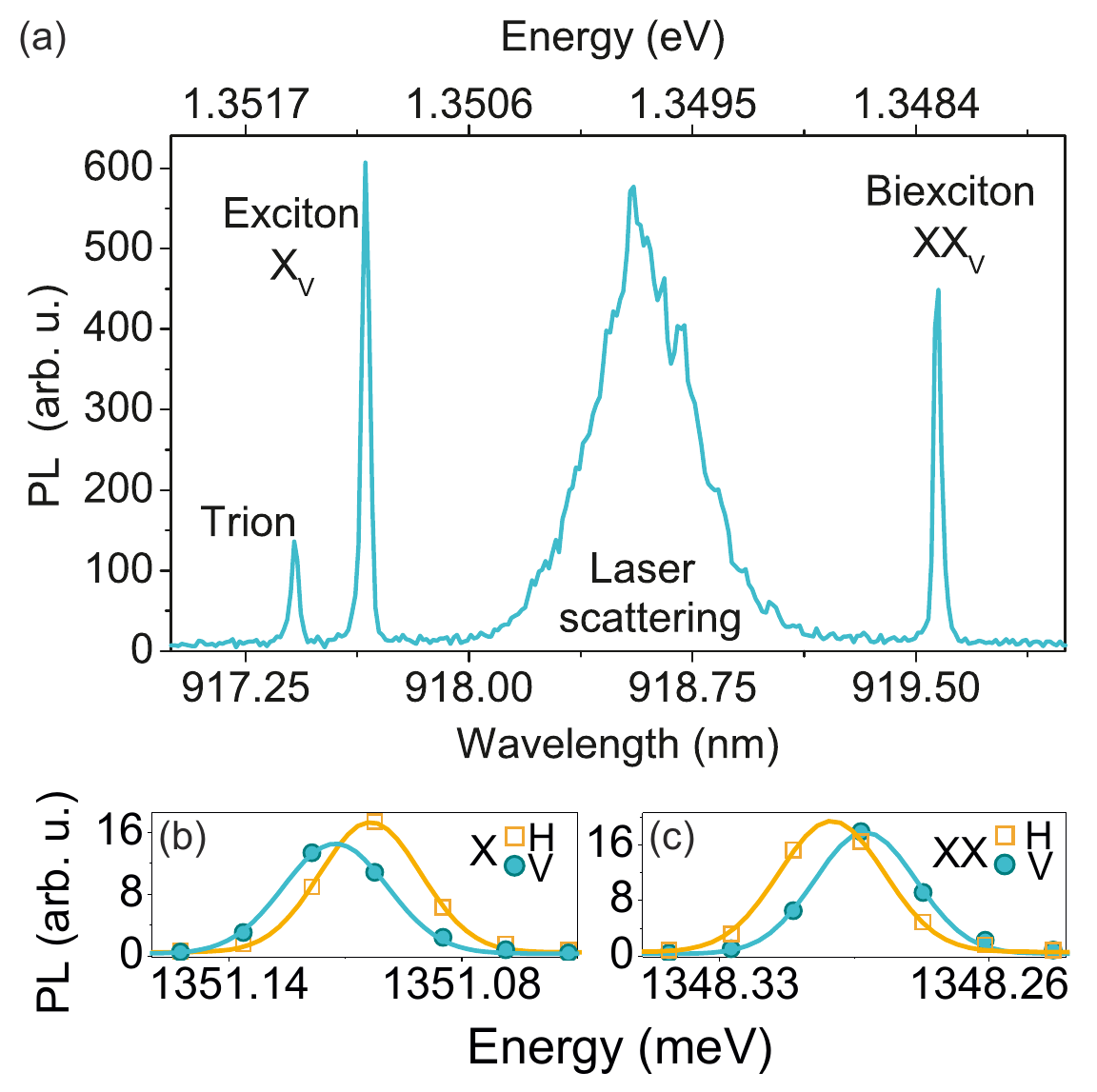}
\caption{ Photoluminescence under resonant two-photon excitation. (a) Photoluminescence spectra of the V-polarized cascade and a suppressed H-polarized excitation laser. (b) (c) Emission from V and H-polarized cascades under H-polarized excitation. Solid lines are Gaussian fits to the data showing the fine structure splitting of the exciton states.}
\end{figure}

Finally the process that can best explain the damping of the Rabi-oscillations in our data originates from competing non-resonant two-photon excitation processes that involve the quantum dot, for example the creation of an electron-hole pair with one carrier trapped in the dot and one free in the host material. Evidence for this explanation comes from observing exciton and biexciton emission with the excitation laser red detuned with respect to the virtual level [Fig.~3(c)].  In agreement with this proposed mechanism of an incoherent two-photon process we find that this emission has a superlinear population power dependence [Fig.~3(b)]. We do not exclude the existence of a two-photon excitation in the surrounding material, but this process would not cause the damping of the Rabi oscillation but rather a background in the photo-luminescence signal. The theoretical curves [Fig.~3(a)] include the dephasing caused by incoherent two-photon processes and fit the data very well for all values of the detuning of the excitation laser to the two-photon resonance.

\begin{figure}[!ht]
\includegraphics[width=0.95\linewidth]{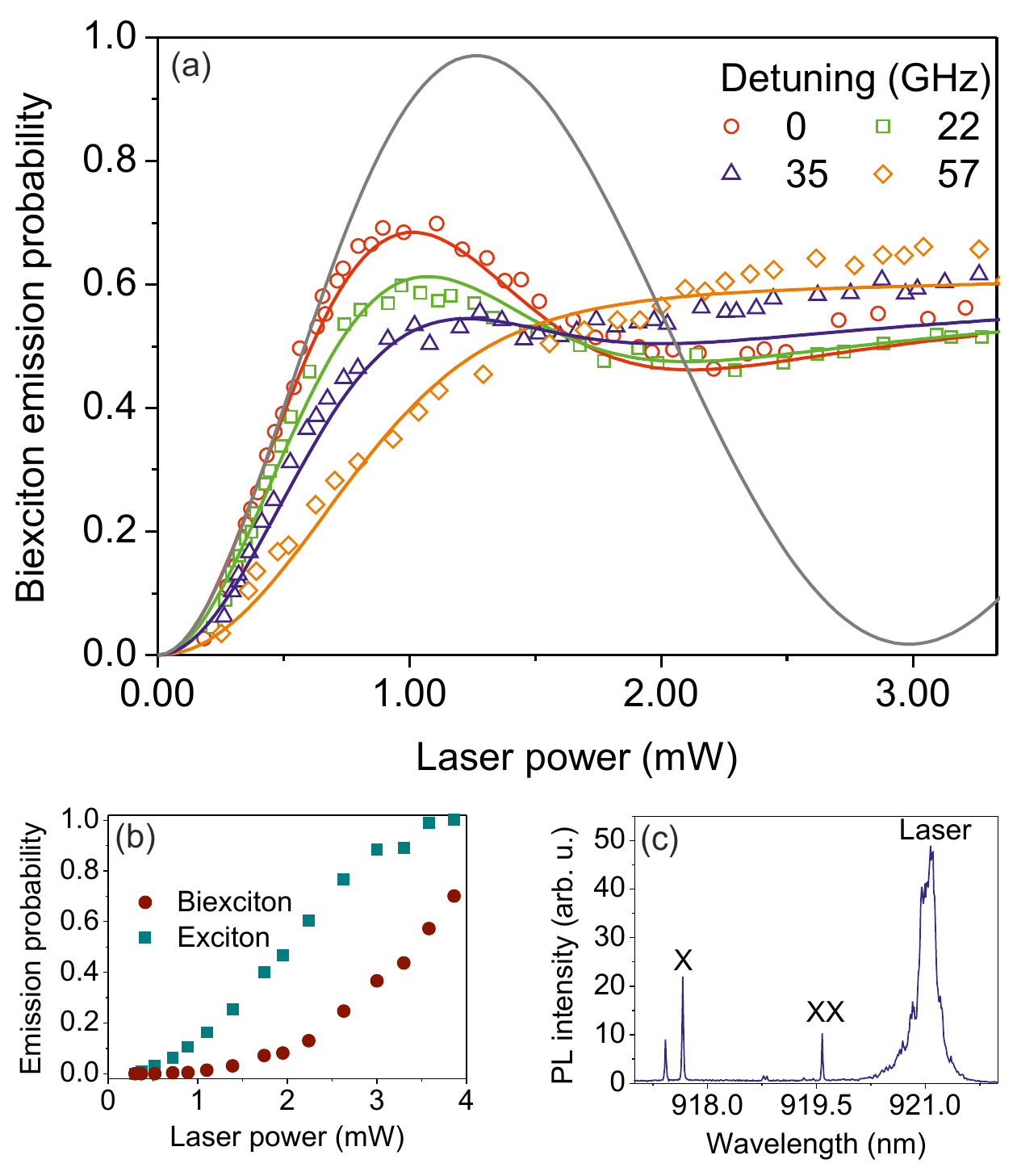}
\caption{ (a) Power dependence of the biexciton emission probability for various excitation laser detunings $(E_{exc} - (|b \rangle-|g \rangle)/2)/h$ from the two-photon resonance. Solid lines are simulations. (b) Power dependence of biexciton and exciton photons under incoherent two-photon excitation, far detuned from two-photon biexciton state resonance towards lower energy. (c) Photoluminescence spectra from the 3mW measurement point of  (b).}
\end{figure}

The resonantly created photon pairs emitted in the cascade allow us, like in the case of spontaneous parametric down-conversion, to use the ratio of the coincidental detections of both cascaded photons to the number of single events to estimate the total detection efficiency $(2.7\permil)$. Using the detection efficiency we obtain a rate of created excitations / photon pairs of 18 MHz at the maximum of the highest Rabi oscillation ($\pi-pulse$).  This reduction in the rate of excitations from the 76 MHz laser repetition rate can be attributed to two sources: decoherence of the excitation process observable in the damping of the Rabi-oscillation and blinking  \cite{Santori} of the quantum dot, which is evident in the auto-correlation measurement through the decreased correlation peaks at long delay times [Fig.~4]. Due to the resonant nature of the excitation process the measurements of the auto-correlations of both exciton and biexciton photons show the full suppression of multi-photon events. With background subtraction the auto-correlations of the V-polarized biexciton and exciton are 0.024(3) and 0.0073(8) at zero delay. For this particular polarization we detected single photons with a rate of 24 kHz in a single mode fiber and photon pairs with a rate of 62 Hz.

\begin{figure}[!ht]
\includegraphics[width=0.95\linewidth]{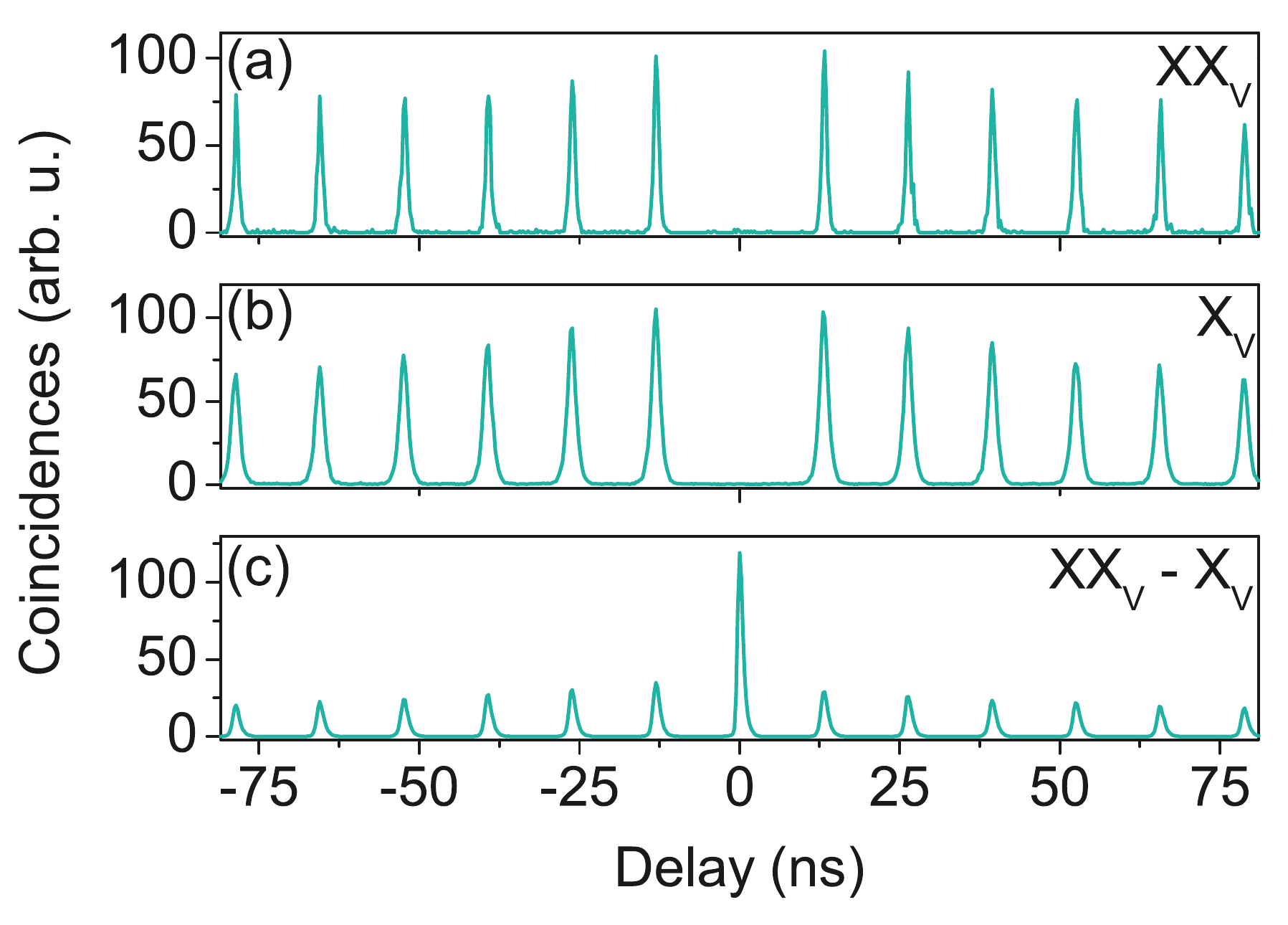}
\caption{(a) (b) Auto-correlation measurement of the V-polarized biexciton and exciton photons in resonant excitation. (c) Cross-correlation coincidence counts between the biexciton (start) and exciton (stop) photons.}
\end{figure}

The coherence of the excitation process enables us to manipulate the phase of the ground-biexciton state superposition. To perform coherent control of this qubit we use a sequence of two consecutive pulses that are derived by feeding pulsed laser light into a Michelson interferometer (pump interferometer in Fig.~1(b)).  With this method we observe Ramsey interference fringes in both exciton (supplementary information) and biexciton [Fig.~5] emission. The power of the first pulse in the Ramsey sequence is adjusted to reach half of the maximal biexciton state population ($\pi/2-pulse$). The second pulse, delayed in time, can either map the population back to the ground state or further increase it to the biexciton state, depending on the relative phase between the pulses. When the time delay between the pulses is shorter than the pulse coherence we observe direct interference of the laser pulses [Fig.~5(b)]; when the delay exceeds the coherence length of the laser we observe interference originating from the atomic superposition. The observed visibility of the Ramsey fringes decays with 179 ps $(T_{2}^*)$ for the biexciton [Fig.~5] and 182 ps $(T_{2}^*)$ for the exciton (supplementary information). 

\begin{figure*}[!ht]
\includegraphics[width=0.95\linewidth]{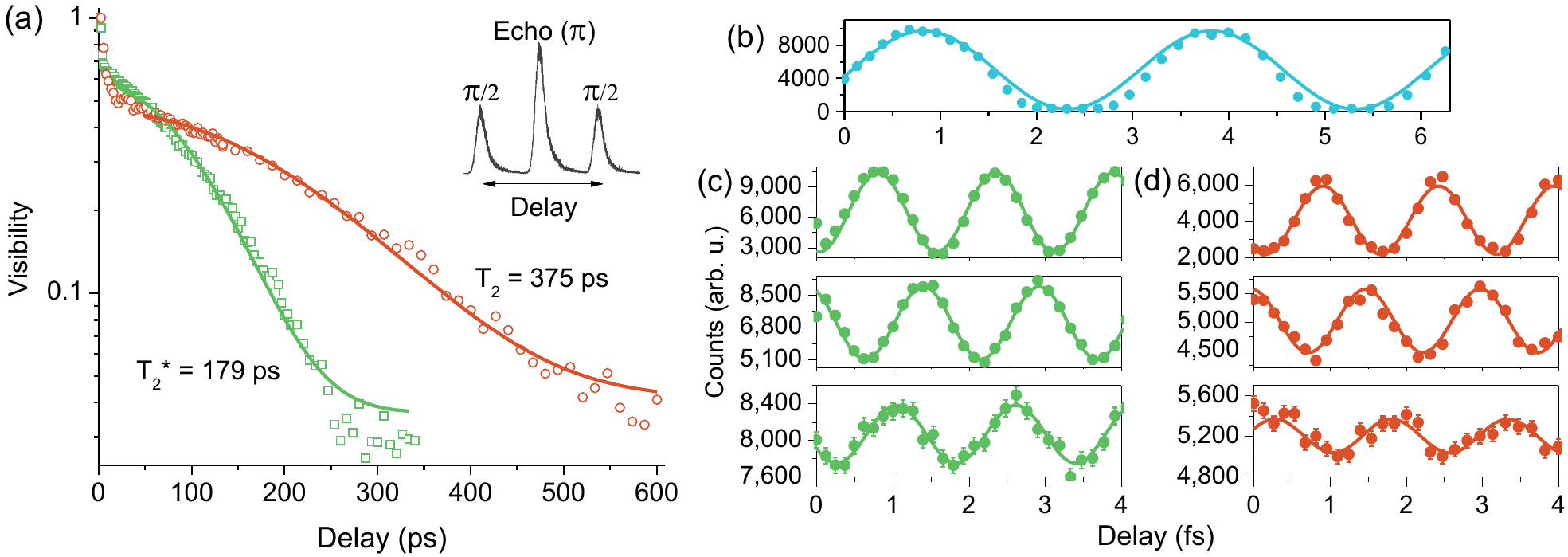}
\caption{(a) Visibility of Ramsey interference in the two – pulse (green open squares) and three – pulse echo sequence (red open circles) coherent control experiment monitored with the biexciton photons. (b) Laser interference at 0 ps coarse delay. (c) Ramsey interference fringes at coarse delays of 8 ps, 107 ps, and 240 ps.  (d) Ramsey interference fringes after the echo pulse at coarse delays of 45 ps, 339 ps, and 472 ps. Both Ramsey and echo sequence measurements do not reach zero visibility due to shot noise of $\sqrt N$ counts. Error bars for the shot noise are visible only for long delays.}
\end{figure*}

A coherent system is Fourier transform limited with $ T_2 = 2T_1$. When compared with the lifetimes $(T_1)$ of the states of 405 ps (biexciton) and 771 ps (exciton) the Ramsey coherence times of the ground-biexciton qubit are shorter, and thus probably affected by phase noise. In addition, the contrast decays as a Gaussian exponential, which further indicates the presence of inhomogeneous broadening.   To investigate the nature of the noise we performed a spin-echo measurement. For this we used the Michelson interferometer in a double-pass configuration capable of delivering the three consecutive pulses necessary for the spin-echo sequence ($\pi/2, \pi, \pi/2$). We observe an increase in the coherence time for biexciton ($T_2 $ = 375 ps) and exciton ($T_2$ = 388 ps). The measured values along with the preserved Gaussian decay indicate the presence of high frequency noise, which could not be refocused by the spin echo technique.  The initial interference visibility in the Ramsey measurement was 0.67 while the same visibility in the echo measurement was 0.4. We attribute this decrease as well as the Ramsey interference not being equal to unity to the damping processes explained above. 

To summarize, we have shown laser-scattering-free emission of coherent photon pairs created by pulsed, resonant, two-photon excitation. Using picosecond optical pulses we performed coherent manipulation of the phase of the ground-biexciton qubit monitored in photoluminescence with a visibility of 67\%. The coherence time of the qubit was more than doubled by an echo sequence. Our photon statistics measurements prove the complete suppression of multi-photon emission events resulting in one of the cleanest single photon sources ever demonstrated. Furthermore, the biexciton photons are created resonantly and the emission process does not involve spontaneous scattering of phonons, which makes the biexciton wave-packet “jitter-free”.   This property increases the indistinguishability \cite{Polyakov} of the biexciton photon, which is an essential condition for the interaction of flying qubits and for linear optical quantum computation schemes. Deterministic creation of single photons and single photon pairs in combination with polarization entanglement  \cite{Stevenson} is a step forward in the realization of quantum information protocols with quantum dots. The possibility to coherently transfer the phase of the excitation light to the excited system makes the presented excitation scheme suitable for the creation of time-bin entanglement from quantum dots  \cite{Simon}.

\begin{acknowledgments}
We acknowledge discussions with H. Ritsch and M. Hennrich regarding the theoretical modeling and with C. Couteau on the early ideas leading to this work. This work was funded by the European Research Council (project EnSeNa) and Canadian Institute for Advanced Research through its Quantum Information Processing program. A.P. would like to thank Austrian Science Fund (FWF) on support provided through Lise Meitner Postdoctoral Fellowship M-1243. G.S.S. acknowledges partial support through the Physics Frontier Center at the Joint Quantum Institute (PFC@JQI).
\end{acknowledgments}

\onecolumngrid
\setstretch{2}
\textbf{Supplementary Information}

\textbf{Materials and Methods:}

 Our sample contained self-assembled InAs quantum dots of low density (approximately $10\mu m^{-2}$) grown by molecular beam epitaxy. The quantum dots were embedded in a $4\lambda$ (542.7 nm of GaAs) distributed Bragg reflector (DBR) microcavity consisting of 15.5 lower and 10 upper $\lambda$/4 thick DBR layer pairs of AlAs and GaAs, with a cavity mode at $\lambda$ = 920 nm. The sample was kept in a helium-flow cryostat temperature stabilized to 5.0 $\pm$ 0.1 K. The excitation pulses were derived from a 76 MHz Ti:Sapphire laser.  To spectrally limit the scattered laser light, the pulse length was conveniently adjusted by a pulse-shaper, which consisted of two diffraction gratings and a slit placed in-between them.  The excitation light was focused onto the sample from the side. Here the sample DBR structure acted as a waveguide for the excitation laser. The emission was collected from the top (orthogonal to the excitation plane) using a 60x microscope objective.  The emission (biexciton and exciton photons) was spectrally separated in a home-built spectrometer and coupled into single mode fibers. We measured the coherence length of the emitted photons. For this purpose we send the collected biexciton (exciton) photons onto a Michelson interferometer, where we observed the interference signal as the length of the interferometer was scanned. 

To create the pulse sequence for coherent manipulation and spin echo measurements we used a Michelson interferometer. It could be used either in a single pass configuration which produced a sequence of two consecutive pulses or in a double pass configuration for which the output sequence had three pulses.  The single pass configuration resulted in two equal-intensity pulses with controllable delay, which we used in coherent control Ramsey measurements ($\pi/2-\pi/2$ sequence). The sequence of three pulses was used for spin echo measurements and here the middle pulse was a result of interference of the first and second interferometer passes. Consequently, using a phase control (a phase plate in the first pass beam path) we could adjust the middle pulse to have twice the intensity of the other pulses in the sequence ($\pi/2-\pi-\pi/2$  sequence). We controlled the coarse delays of the interferometer by a motorized linear stage and fine adjustments by a piezoelectric transducer.

\textbf{Theory:}

The levels involved are the ground ($├ |g\rangle$), exciton state ($├ |x\rangle$), and biexciton state ($├ |b\rangle$).  The level scheme is shown in Fig.~1(a) of the main text. The energy differences between ground state and exciton state, and between exciton state and biexciton state are not equal due to the increased biexciton binding energy $E_{b}$ with respect to the single exciton. This electronic configuration allows for a two-photon excitation process where the pump laser is not resonant to any of the single photon transitions, while the two-photon process is resonant. The Hamiltonian used to describe this system is given as:
\begin{equation}
H=\frac{\hbar\text{\ensuremath{\Omega_{1}\left(t\right)}}}{2}(\sigma_{gx}+\sigma_{gx}^{\dagger})+\frac{\hbar\text{\ensuremath{\Omega_{2}\left(t\right)}}}{2}(\sigma_{bx}+\sigma_{bx}^{\dagger})+\hbar\sigma_{xx}(\text{\ensuremath{\Delta_{e}}}-\text{\ensuremath{\Delta_{b}}})-2\hbar\sigma_{bb}\text{\ensuremath{\Delta_{b}}}
\end{equation}
Here, $\Omega_{l}(t)$ is the Rabi frequency of the pump laser driving both single photon transitions. The transition operators or projectors are given as $\sigma_{(i,j)}=|i\rangle\langle j┤| $ .  The detuning $ \Delta_{e}=E_{b}/(2\hbar)=2\pi*335$ GHz between the two-photon transition virtual level and the exciton energy. If we want to drive the two-photon transition off-resonantly we define the detuning ($\Delta_{b}$), the difference between the two-photon resonance and the driving laser energy.
\begin{equation}
H=\hbar\left(\begin{array}{ccc}
0 & \frac{\text{\ensuremath{\Omega_{1}\left(t\right)}}}{2} & 0\\
\frac{\text{\ensuremath{\Omega_{1}\left(t\right)}}}{2} & -\text{\ensuremath{\Delta_{b}}}+\text{\ensuremath{\Delta_{e}}} & \frac{\text{\ensuremath{\Omega_{2}\left(t\right)}}}{2}\\
0 & \frac{\text{\ensuremath{\Omega_{2}\left(t\right)}}}{2} & -2\text{\ensuremath{\Delta_{b}}}
\end{array}\right).
\end{equation}
For writing the master equation in Lindblad form,
\begin{equation}
\dot{\rho}=-\frac{i}{\hbar}[H,\rho]+\sum_{i=1}^{4}\mathcal{L}_{i}(\rho)
\end{equation}
we use the following Lindblad operators:
\begin{equation}
\mathcal{L}_{1}(\rho)=\frac{\text{\ensuremath{\gamma_{b}}}}{2}(2\sigma_{bx}^{\dagger}\rho\text{\ensuremath{\sigma_{bx}}}-\sigma_{bx}\sigma_{bx}^{\dagger}\rho-\rho\sigma_{bx}\sigma_{bx}^{\dagger})
\end{equation}
\begin{equation}
\mathcal{L}_{2}(\rho)=\frac{\gamma_{x}}{2}(2\sigma_{gx}\rho\sigma_{gx}^{\dagger}-\sigma_{gx}^{\dagger}\sigma_{gx}\rho-\rho\sigma_{gx}^{\dagger}\sigma_{gx})
\end{equation}
\begin{equation}
\mathcal{L}_{3}(\rho)=\frac{\gamma_{dephb}}{2}(2(\sigma_{bb}-\sigma_{xx})\rho(\sigma_{bb}-\sigma_{xx})^{\dagger}-\rho(\sigma_{bb}-\sigma_{xx})^{\dagger}(\sigma_{bb}-\sigma_{xx})-(\sigma_{bb}-\sigma_{xx})^{\dagger}(\sigma_{bb}-\text{\ensuremath{\sigma_{xx}}})\rho)
\end{equation}
\begin{equation}
\mathcal{L}_{4}(\rho)=\frac{\gamma_{dephe}}{2}(2(\sigma_{xx}-\sigma_{gg})\rho(\sigma_{xx}-\sigma_{gg})^{\dagger}-\rho(\sigma_{xx}-\sigma_{gg})^{\dagger}(\sigma_{xx}-\sigma_{gg})-(\sigma_{xx}-\sigma_{gg})^{\dagger}(\sigma_{xx}-\sigma_{gg})\rho)
\end{equation}
Here, $\gamma_{b}$ and $\gamma_{x}$ are the spontaneous decay rates and $\gamma_{db}$  and $ \gamma_{dx}$  are the dephasing rates of the biexciton and exciton, respectively. The excitation pulse is considered to be a Gaussian of the following form
\begin{equation}
\ensuremath{\Omega_{l}[t,\Omega,\sigma]=\Omega \cdot exp[-(2 log(2)\cdot t^2)/\sigma^2 ]}
\end{equation}
where $\Omega$ is the maximum Rabi frequency of the laser pulse and $\sigma$=4 ps is the pulse duration.

Parameters like spontaneous decay and dephasing rates were determined from experimental lifetime and coherence time measurements, respectively. Using these experimentally obtained parameters we can numerically solve the master equation. For our experimental system the measured lifetimes are $\tau_{b}=1/\gamma_{b} $ = 405 ps for the biexciton and$\tau_{x}=1/\gamma_{x} $ = 771 ps for the exciton. The coherence lengths are $\tau_{db}=1/\gamma_{db}$  = 211 ps for the biexciton photon and $\tau_{dx}=1/\gamma_{dx}$ = 119 ps for the exciton photon, measured interferometrically. 
By solving the master equation for the given experimental parameters we obtain the populations of the different levels involved $(P_{i}=\langle \sigma_{ii} \rangle)$. Population integrated over time gives the emission probability. The emission probability as a function of the square of Rabi frequency in resonant excitation shows an oscillating behavior commonly known as Rabi oscillations. The result of this simulation is given in Fig. 3(a) of the main text as a grey curve.
A simple way to model the influence of a process such as incoherent two-photon generation of excitons and biexcitons is to add two more Lindblad terms that drive the population incoherently from the ground state to the exciton state and from the exciton state to the biexciton state, respectively. The rates for these processes are proportional to the Rabi frequency to the power of four since they are based on a two two-photon excitation processes.
\begin{equation}
\text{\ensuremath{\mathcal{L}_{5}}(\ensuremath{\rho})}=-\frac{\Omega_{3}^{4}}{2}(\sigma_{bx}^{\dagger}\sigma_{bx}\rho+\rho\sigma_{bx}^{\dagger}\sigma_{bx}-2\sigma_{bx}\rho\sigma_{bx}^{\dagger})
\end{equation}
\begin{equation}
\text{\ensuremath{\mathcal{L}_{6}}(\ensuremath{\rho})}=-\frac{\Omega_{3}^{4}}{2}(\sigma_{gx}\sigma_{gx}^{\dagger}\rho+\rho\sigma_{gx}\sigma_{gx}^{\dagger}-2\sigma_{gx}^{\dagger}\rho\sigma_{gx})
\end{equation}
Solving the master equation (3) including these two additional terms give us the theoretical curves which are depicted in Fig.~3(a) of the main text.  These curves were calculated using the above experimental parameters and the four two-photon detunings $\Delta_{b}$ =$2\pi \{ 0, 22, 35, 57 \}$ GHz. The only free parameter of the fit was the relative strength of the incoherent process expressed through a constant k in $\Omega_{i} (t)=k\Omega_{l}(t)$. This constant was found to be k=0.47 and is the same for all four theory curves. The experimental data was normalized to the theoretical values by a common numerical factor. The results of the power dependence measurement of the incoherently created emission were also scaled with the same scaling factor. This gave a unit emission probability at the saturation of the incoherent excitation, which further confirms the correctness of the model. 
The emission probability and the detection efficiency were estimated in the following way: the single count rates are 23 kHz and 24 kHz for biexciton and exciton, respectively, while the coincidence count rate is 62 Hz for detection made with cross-polarizers. When measured without this scattering suppression technique these rates double. This yields heralding efficiency of $\eta$=2.7$\permil$ which is used to determine the losses as 1-$\eta$. Scaling the number of detected single counts to the losses we come to a number close to 18 MHz.
Without subtraction of the background caused from the coincidence events between signal and the detector dark counts the auto-correlation parameters are 0.012(1) for exciton and 0.0314(4) for biexciton.
\begin{figure}[!ht]
\includegraphics[width=0.95\linewidth]{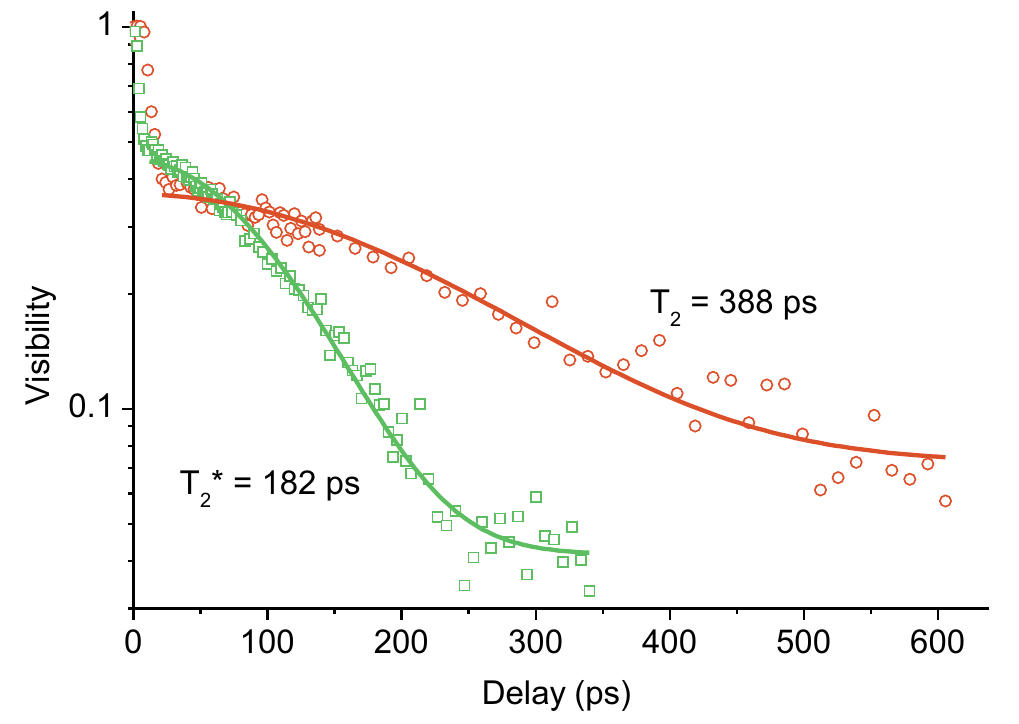}
\caption{ Coherent manipulation monitored with the exciton photons.  Visibility of Ramsey interference in the two-pulse (green open squares) and three-pulse echo sequence (red open circles) coherent control experiment.}
\end{figure}

\end{document}